\documentclass[aps,prb,reprint,preprintnumbers,amsmath,amssymb,floatfix,longbibliography]{revtex4-1}
\usepackage{graphicx}
\usepackage{dcolumn}
\usepackage{bm}
\usepackage{color}
\usepackage{tabularx}
\usepackage{amsmath,amsfonts,amsthm,amssymb}
\usepackage{appendix}
\usepackage[bookmarksnumbered,pdfpagelabels=true,plainpages=false,colorlinks=true,linkcolor=blue,citecolor=blue,urlcolor=blue]{hyperref}
\usepackage[english]{babel}

\newcommand{\cD}{\mathcal{D}}
\newcommand{\cM}{\mathcal{M}}
\newcommand{\vS}{\mathbf{S}}
\newcommand{\vD}{\mathbf{D}}
\newcommand{\vv}{\mathbf{v}}
\newcommand{\vnabla}{\mathbf{\nabla}}
\newcommand{\vF}{\mathbf{F}}

\newcommand{\vB}{\mathbf{B}}

\begin{document}
\title{Motion-induced inertial effects and topological phase transitions in skyrmion transport}
\author{A. W. Teixeira$^{1}$}
\author{S. Castillo-Sep\'ulveda$^{2}$}
\author{L. G. Rizzi$^{1}$}
\author{A. S. Nunez$^{3}$}
\author{\\R. E. Troncoso$^{4}$}
\author{D. Altbir$^{5}$}
\author{J. M. Fonseca$^{1}$}
\author{V. L. Carvalho-Santos$^{1}$}
\affiliation{$^{1}$Departamento de F\'isica, Universidade Federal de Vi\c cosa, 36570-900, Vi\c cosa, Brazil}
\affiliation{$^{2}$Departamento de Ingenier\'ia, Universidad Aut\'onoma de Chile, Providencia, Chile}
\affiliation{$^{3}$Departamento de F\'isica, FCFM, CEDENNA, Universidad de Chile, Santiago, Chile}
\affiliation{$^{4}$Center for Quantum Spintronics, Department of Physics, Norwegian University of Science and Technology, NO-7491 Trondheim, Norway}
\affiliation{$^{5}$Departamento de F\'isica, CEDENNA, Universidad de Santiago de Chile, Santiago, Chile}

\date{\today}

\begin{abstract}
In this work, the current-induced inertial effects on skyrmions hosted in ferromagnetic systems are studied. {When the dynamics is considered beyond the particle-like description, magnetic skyrmions can deform due to a self-induced field. We perform Monte Carlo simulations to characterize the deformation of the skyrmion during its movement}. In the low-velocity regime, the deformation in the skyrmion shape is quantified by an effective inertial mass, which is related to the dissipative force. When skyrmions move faster, the large self-induced deformation triggers topological transitions. The transition is characterized by the proliferation of skyrmions and different total topological charge, which are obtained in terms of the skyrmion velocity. Our findings provide an alternative way to describe the skyrmion dynamics that take into account the deformations of its structure. Furthermore, the motion-induced topological phase transition brings the possibility to control the number of ferromagnetic skyrmions by velocity effects.
\end{abstract}

\maketitle

\section{Introduction}
Over the last decades physical systems with topological protection have been the focus of extensive research \cite{P1,P2,P3,P4}.  Particle-like excitations, characterized by non-trivial topological invariants, cannot be continuously deformed to another state with different topology, unless enough extra energy is injected into the system. Magnetic skyrmions \cite{Skyrme,BP} are an example of topologically stable spin structures. These textures, characterized by a topological charge, have been predicted \cite{Nagaosa,Jonietz,Munzer,Milde,Nagao,Seki,Muhlbauer} in magnetic systems with Dzyaloshinkii-Moriya interaction (DMI) \cite{Dzyaloshinskii,Moriya}, known as chiral magnets. The observation of magnetic skyrmions, in non-centrosymmetric crystals \cite{Muhlbauer,Pappas,Tonomura,Moskvin}, cubic helimagnets \cite{Yu465,Yu10} and ultrathin films \cite{Heinze,Romming}, opened the door to potential functionalities, e.g., as information bits in spin-based devices \cite{Parkin,Kiselev,Tomasello,Fert} or neuromorphic systems \cite{Ehrmann,Grollier1,Grollier2}. Although it is known that applications of skyrmions in spintronics demand specific conditions, such as low current densities or temperature gradients \cite{Fert,Zang,Sampaio,Iwasaki8,Iwasaki13,Iwasaki14,Troncoso,Schutte,Kong,Kovalev,Lin,Knoester,Yu988}, the {difficulties to generate skyrmions and displace them straight along the applied currents}  \cite{Iwasaki13,Hall-1}  have  hampered  the  progress  of skyrmion-based spintronics.

The dynamics of magnetic skyrmions is usually described by the Thiele's equation \cite{Thiele}, which derives from the Landau-Lifshitz-Gilbert (LLG) equation \cite{Landau,Gilbert,Everschor,Liu,Nagaosa} under the assumption of a rigid structure. However, if the rigidity condition is relaxed, one allows the skyrmion shape to be deformed and the texture becomes dependent on its velocity. As a result, the effective dynamics is now captured by an inertial Thiele's equation \cite{Schutte,Troncoso-AP}, where the mass appears as the capacity of the system to store kinetic energy during the motion \cite{Makhfudz,Moutafis,Troncoso-AP}. Massive dynamics of skyrmions has been previously studied \cite{Schutte,Makhfudz} and observed in magnetic CoB/Pt \cite{Buettner} and FeGe disks \cite{ZhaoPNAS}. {Recently, it has been shown \cite{LiuPRA} that beyond the small-driving-current regime, deformation of skyrmions consisting of an expanding size and noncircular shape, occurs during the transport. The last yields interesting nonlinear effects in the relation between the charge-current and skyrmion velocity.} 
 
Topological transitions have been predicted for certain types of magnetic textures as merons \cite{G1,G2}, i.e., a magnetic vortex with a core. It was shown that under large applied charge currents, the ultrafast switching of the vortex polarity, generates a magnetic singularity during the vortex annihilation \cite{Nature-2006,Thiaville-PRB,Gabriel-Work}. This process, mediated by the appearing of an unstable antivortex, induces a change in the topological charge, $q=p\, w$, with $p$ the vortex polarity and $w$ the chirality. Topological transitions in skyrmion systems have been recently observed in magnetic multilayers\cite{Arxiv-2020}, where the number of skyrmions are thermally controlled. In spite of some efforts \cite{Erica}, a systematic study of topological  transitions of skyrmions induced by magnetic fields, temperature or charge currents has not been addressed.

In this paper, we study the self-induced deformation of moving skyrmions at low temperature. Using Monte Carlo simulations we characterize the shape of current-induced skyrmions in terms of its velocity. We have analyzed two main regimes to describe the skyrmion profile. At low velocities, the current-induced motion produces an effective field that modifies slightly the shape of the skyrmion \cite{Troncoso-AP}. The later is quantified by a massive term, which is determined as a function of the velocity and found that is approximately constant for very low speeds. For large velocities, we show that the deformation of skyrmions are abrupt and generates transitions that change the total topological charge.
We find that at certain critical velocities, the self-induced field provides enough magnetic energy to deform the skyrmion texture by increasing its topological charge.

This paper is organized as follows. In Sec. \ref{ThModel} we present the theoretical model and details of the Monte Carlo simulations. Sec. \ref{results} presents our results and related discussion. In Sec. \ref{Conclusions} we summarize the conclusions and prospects.

\section{Model and numerical methods}\label{ThModel}
\subsection{Classical spin Hamiltonian}
\begin{figure}[!]
    \centering
    \includegraphics[scale=0.45]{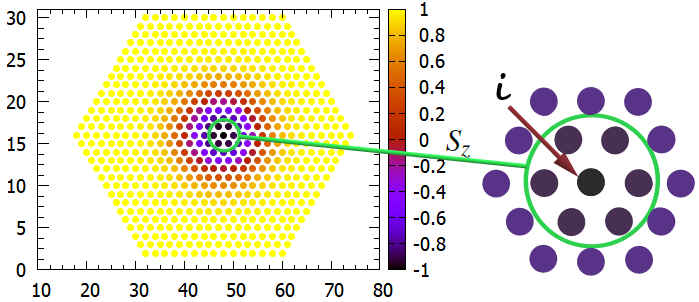}
    \caption{Representation of a magnetic skyrmion centered at the hexagonal spin lattice with $631$ sites. The color code indicates the $S_z$ component of the magnetization. The inset indicates the $i-${th} lattice site and its nearest neighbors.} 
    \label{SpinLattice}
\end{figure}

We consider a ferromagnet on a hexagonal lattice, as shown in Fig. \ref{SpinLattice}. The spin system is described by the Hamiltonian,
\begin{eqnarray}\label{hamiltonian}
H= &-&J\sum_{\langle i,j\rangle} \vS_{i} \cdot \vS_{j} - \mu \sum_{i}\ \vB \cdot \vS_{i} - K\sum_{i}\ S_{iz}^2 
\nonumber \\
&-& \sum_{\langle i,j\rangle} \vD_{ij}\cdot \left(\vS_{i}\times \vS_{j}\right)\,,
\label{InitialHamiltonian}
\end{eqnarray}
where $\langle ,\rangle$ stands for the summation over nearest neighbors (NN) lattice sites. In Eq. (\ref{InitialHamiltonian}), $J > 0$ is the ferromagnetic exchange coupling constant between NN spins, $K$ is the easy-axis magnetic anisotropy constant and $\vB$ represents the applied magnetic field along the $x$-direction. In addition, a Dzyaloshinskii-Moriya interaction (DMI) is included to stabilize the magnetic skyrmion. The DMI favours a canted orientation between NN spins and it is characterized by the vector $\vD_{ij}=D {\bf d}_{ij}$, where $D$ is the DMI strength. The pattern of the magnetization vector depends on whether ${\bf d}_{ij}$ is perpendicular or parallel to the vector $\mathbf{r}_{ij}$ that connects two neighboring spins. In this work, we consider $\mathbf{r}_{ij}\cdot{\bf d}_{ij}=0$, so that Bloch-type skyrmions will be stabilized.

\subsection{Self-induced field by the skyrmion motion}\label{SIEFSM}
Let us consider the dynamics of single skyrmions induced by electrical currents. From the LLG equation one has that the skyrmion texture self-induces a field \cite{Troncoso-AP},
\begin{equation}\label{eq:Bsi}
\vB_{si} = \frac{2\pi}{\gamma a}\vS\times\left(\vv\cdot \vnabla\right)\vS\,,
\end{equation}
which depends linearly on the skyrmion velocity $\vv$ and its magnetization $\vS$. In the previous expression $\gamma$ is the gyromagnetic ratio and $a$ the lattice constant. Thus, as the skyrmion moves, the field ${\bf B}_{si}$ modifies the shape of its spin texture. The magnetization pattern changes gradually until the system reaches the steady-state, resulting in an asymmetrically deformed skyrmion.

Representing the magnetization profile by the ansatz ${\bf S}({\bf r},t)={\bf S}({\bf r}-{\bf x}(t),{\bf v}(t))$, the deformed skyrmion dynamics can be described through the generalized Thiele's equation \cite{Schutte,Troncoso-AP},
\begin{equation}\label{ThieleEq}
{\cal M}\dot{\vv}(t) +g\hat{z}\times\vv(t)+\alpha\cD\vv(t) = \vF \; ,
\end{equation}
where the external force is $\vF$, while the dissipative force is proportional to the Gilbert damping $\alpha$ and the elements $\cD_{mn}=\int d^2x \left({\partial_{m}\vS} \cdot {\partial_{n} \vS}\right)$ with $({m,n}= {x,y})$. Note that for skyrmions at rest, the matrix elements of $\cD$ satisfy $\cD_{xx}=\cD_{yy}\equiv \cD_r$. The Magnus force is proportional to
$g=4\pi Q$, where $Q= ({4\pi})^{-1}\int d^2x\ \vS\cdot \left({\partial_x\vS} \times {\partial_y \vS}
\right)$ is the skyrmion charge. The effective mass in the inertial term in Eq. (\ref{ThieleEq}) is described by the matrix ${\cal M}={\cal M}^c+{\cal M}^d$, where ${\cal M}^c_{mn}=\int d^2x \vS\cdot\left({\partial_{m}}\vS \times {\partial_{v_n} \vS}\right)$ and ${\cal M}^d_{mn}=\alpha\int d^2x \left({\partial_{m}}\vS \cdot {\partial_{v_n} \vS}\right)$, which are originated both from the conservative and dissipative dynamics, respectively.

In the limit of low velocities, the magnetization texture of a non-rigid steady-state moving skyrmion can be represented perturbatively as \cite{Troncoso-AP} ${\bf S}({\bf r},t)={\bf S}_0({\bf r})+\lambda\xi{\bf S}_0({\bf r})\times\left({\bf v}\cdot\nabla\right){\bf S}_0({\bf r})$, with ${\bf S}_0({\bf r})$ being the magnetization texture of the static skyrmion. The dimensionless factor $\lambda$ determines the strength at which the skyrmion is deformed and $\xi=\hbar \ell^2/Ja^2$, where $\ell$ is the skyrmion size. Using this approximation, the elements of the mass tensor ${\cal M}$ can be readily obtained from the dissipative term in Thiele's equation and the skyrmion charge $Q$. In particular, the diagonal elements describe the effective scalar mass and satisfy $\cM_{xx}=\cM_{yy}=\lambda\xi\cD_{xx}$, while the non-diagonal terms are given by \cite{Troncoso,Troncoso-AP,Martinez} $\cM_{xy}=-\cM_{yx} = 4\pi \lambda \xi \alpha Q$. In addition, since $Q$ is invariant under smooth deformations, the non-diagonal elements of the mass matrix do not change when skyrmions moves slowly. However, for high velocities the shape deformation is large enough and, as we will show later, a topological transition may occur.

\subsection{Monte Carlo simulations}\label{MCSimu}
In order to investigate the steady-state profile  of  moving  skyrmions  we  use standard Monte Carlo simulations. Based on the Metropolis algorithm \cite{Landau-book} the stable magnetic state for the Hamiltonian (\ref{hamiltonian}) is determined. The temperature of the system $T$ is assumed to be low enough in order to obtain an equilibrium magnetization configuration, that results in a Bloch-type skyrmion. The simulation is performed as follows: for a specific site we propose a random change in the orientation of its  magnetic moment. The criteria to accept changes in the orientation of the respective single magnetic moment is given by the acceptance probability $p=\exp\left[-\Delta H/k_B T\right]$, where $\Delta H$ is the change in the energy of the system. If $p< r$, where $r$ is a random number in the range $[0,1]$, the reorientation is not accepted. One Monte Carlo step (MCS) is defined by $N$ single-site attempts to change the orientation of magnetic spins at different lattice sites.

The effects of the skyrmion velocity can be taken into account by considering the presence of the  self-induced field ${\bf B}_{si}$. For simplicity, we assume a constant velocity along the $x$-direction, $\vv=v\hat{x}$. The field $\vB_{si}$, which is computed from Eq. (\ref{eq:Bsi}) is included in  the  Hamiltonian (\ref{hamiltonian}), in addition to $\vB$. This new magnetization pattern, in turn, creates a new self-induced field $\vB_{si}$, thereby generating a new skyrmion texture $\vS(\mathbf{r},t)$. 
\begin{figure}[h!]
\begin{center}
\includegraphics[width = 8.5cm]{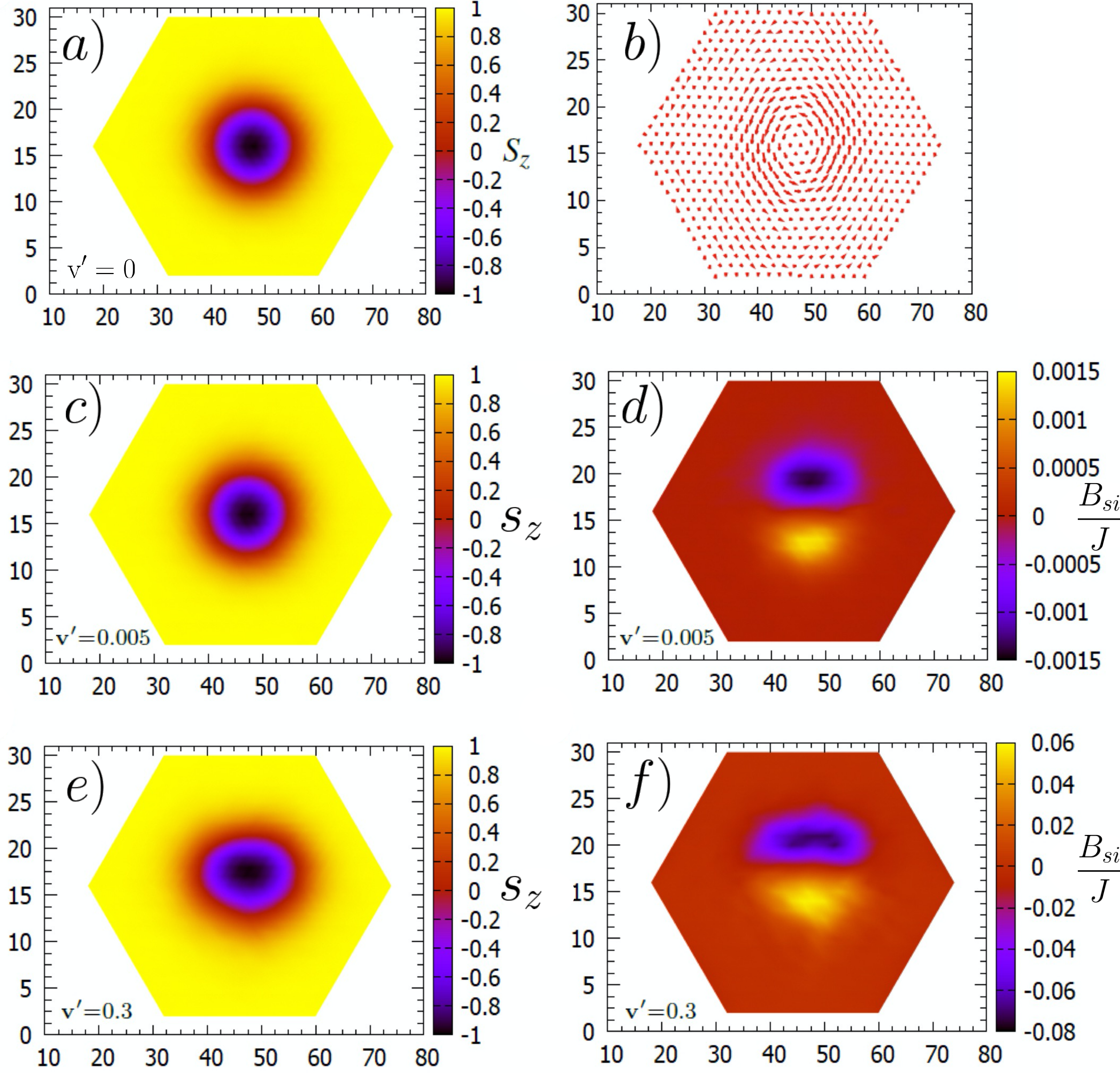}
\caption{(a) $S_z$ component and (b) vector field ($S_x$,$S_y$) of the magnetization for a skyrmion at rest. (c) $S_z$ component of the deformed skyrmion profile and (d) the corresponding $z$-component of the self-induced field $|{\bf B}_{si}|$ of a configuration obtained from MC simulations at a low  velocity $\text{v}'=0.005$. At panels (e)  and  (f) we plot for a velocity $\text{v}'=0.3$ the same quantities as (c) and (d), respectively. Note the asymmetry in the skyrmion shape for the later case.}
\label{fig2}
\end{center}
\end{figure}
This procedure is repeated until the skyrmion profile is stabilized. Once we obtain the stable configuration for the deformed skyrmion, the dissipative matrix $\cD$ is determined and, consequently, the skyrmion effective mass $\mathcal{M}$ is obtained for different velocities. Additionally, we calculate the total topological charge $Q$ of the system to characterize the existing topological transitions.

\section{Results}\label{results}
We consider a system with $631$ Heisenberg spins on a two-dimensional triangular lattice with a hexagonal boundary shape (see Fig. \ref{SpinLattice}) using helical boundary conditions \cite{Hagemeister,Newman}. This system yields a configuration with 15 particles at the edge of the lattice. The used magnetic parameters are defined in terms of $J$, i.e., an external out-of-plane magnetic field $\mu B/J=0.1$, a DMI and anisotropy constants given by $D/J=0.32$ and $K/J=0.07$, respectively. The stable states are obtained for a thermal energy $k_B T/J=0.001$, that corresponds to an absolute temperature of $\sim 80$ mK. As discussed in Ref. [\onlinecite{Hagemeister}], these parameters are adequate to generate a stable skyrmion in a hexagonal lattice. For each set of parameters, we performed $5 \times 10^5$ MCS. To determine the field $\mathbf{B}_{si}$, it is convenient to introduce a dimensionless velocity, $\vv' = ({2\pi\mu}/{J\gamma a})\vv$. In the simulations we consider values for the parameters which are  similar  to  those  experimentally estimated \cite{Hagemeister,Romming} for Pd/Fe/Ir(111), given by  $\mu=3\mu_B$, $J=7\ {\rm meV}$, $a = 2.7\times10^{-10}\ {\rm m}$ and $\gamma=g_L|\mu_B|/\hbar$, where $a$ is the lattice parameter, $\gamma$ the gyromagnetic ratio and $\mu_B$ the Bohr magneton. We  first validate our numerical approach by performing simulations for a skyrmion at rest, i.e., $|\vB_{si}|=0$. As expected, the magnetic configuration of the system consists of a circularly symmetric skyrmion, as shown in Fig. \ref{fig2}(a) and (b).

\subsection{Skyrmion deformation and inertial mass}
Now, we focus on the effects originated from the self-induced field $\mathbf{B}_{si}$ in the skyrmion profile. The initial configuration for the simulations with $|\vB_{si}|\neq0$ consists of a skyrmion pattern obtained for $\text{v}=0$. For small velocities, $\vv'=0.005\, \hat{x}$ ($\vv= 1.5\ {\rm m/s\,}\hat{x}$), the $z$ component of the resulting skyrmion profile is displayed in Fig. \ref{fig2}(c). It can be noticed that the deformation in the shape is small and thus, the profile is almost axially symmetric. The last is consistent with the low values obtained for the field $|{\bf B}_{si}| \approx 0.001\,J$ (see Fig. \ref{fig2}(d)). These are noticeably smaller than other relevant magnitudes of the system and therefore it is expected that the skyrmion mass remains almost constant up to a determined value of velocity. This effect will be evident subsequently, when the dissipative force term is calculated. Larger values of the skyrmion velocity lead to significant changes in its shape. Indeed, as is shown in Fig. \ref{fig2}(e), the magnetization profile is clearly deformed for $\vv'=0.3 \hat{x}$.
The shape deformation is also confirmed by the calculation of the field ${\bf B}_{si}$ at Fig. \ref{fig2}(f).

\begin{figure}[h!]
\begin{center}
\includegraphics[width = 8.7cm]{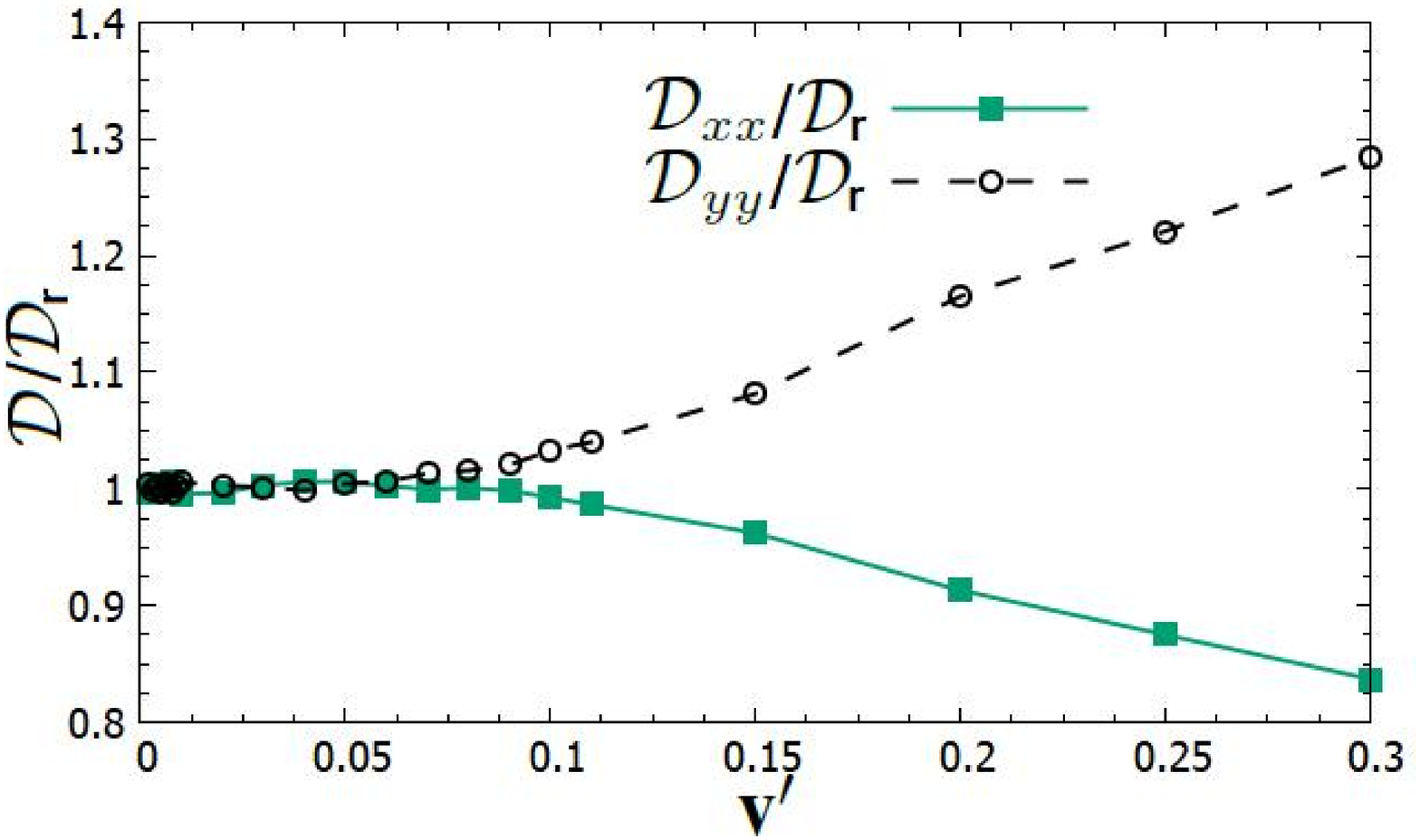}
\caption{Diagonal elements of the dissipative tensor $\cD$ as a function of velocity. The result for $\cD_{xx}$ (green) and $\cD_{yy}$ (black) is normalized by $\cD_r$, i.e., the obtained value for a skyrmion at rest.}
\label{fig3}
\end{center}
\end{figure}

Since the increase in the skyrmion velocity leads to larger changes in its shape, we now study the effects on the effective mass. As discussed in Sec. \ref{SIEFSM} (and also in Ref. [\onlinecite{Troncoso}]), the mass of skyrmions can be directly related to the diagonal elements of the matrix ${\cal D}$. The results depicted in Fig. (\ref{fig3}) indicate that the normalized dissipative parameters $\cD_{xx}/\cD_r$ and $\cD_{yy}/\cD_r$ remain almost constant ($\cD_r\approx 4.78\pi$) up to $\text{v}'\approx 0.09$. When $\text{v}'\geq 0.1$, a clear split between $\cD_{xx}$ and $\cD_{yy}$ is observed. A linear growing (decreasing) of $\cD_{yy}(\cD_{xx})$ is obtained as a function of the skyrmion velocity. This result indicates that a higher anisotropic resistance to the motion occurs as $\text{v}'$ increases. Thus, the higher effective mass will be in the direction perpendicular to the velocity of the skyrmion. The behavior in the skyrmion mass is directly linked to the increase of driving forces and skyrmion mobility under high-driven-currents \cite{LiuPRA}.

\subsection{Motion-induced topological phase transitions}

\begin{figure}
\begin{center}
\includegraphics[width = 8.5 cm]{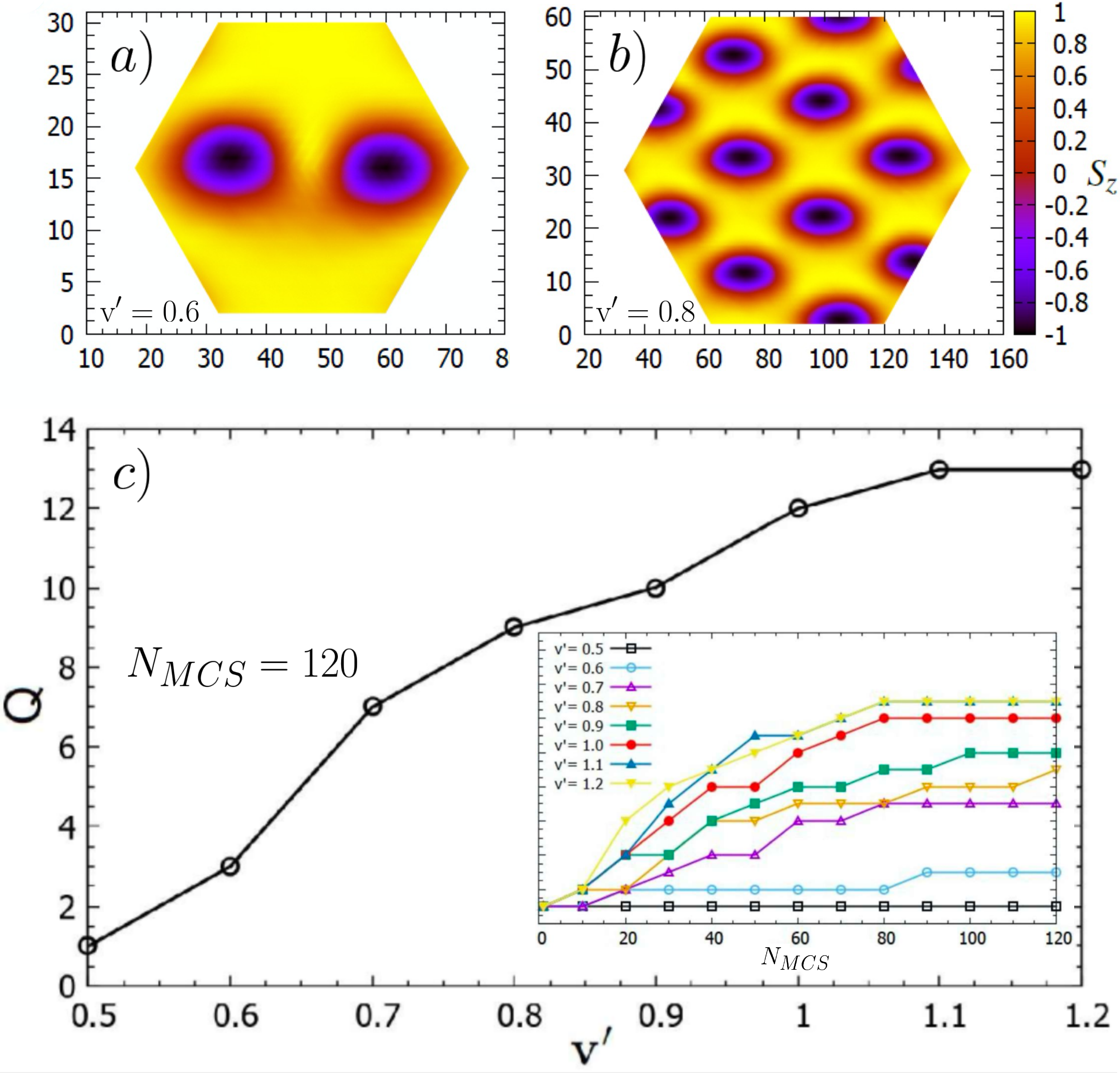}
\caption{{(a) $S_z$ component of the magnetization profile of a system with 15 sites per edge, at $\text{v'}= 0.6$. (b) $S_z$ component of the magnetization profile of a system with 30 sites per edge, at $\text{v'}= 0.8$. (c) Topological charge of the system as a function of the skyrmion velocity. At the inset we display the topological charge in terms of $N_\text{MCS}$ for various velocities.}}
\label{fig5}
\end{center}
\end{figure}

In this section we investigate deformations in the skyrmion texture with focus on the large velocities regime. The simulations reveal that for velocities up to $\text{v}'=0.5$ the skyrmion shape is heavily distorted, without changing the topological charge ($Q=1$). However, for velocities in the range $0.5\leqslant \text{v}'\leqslant0.6$ the number of skyrmions changes. This effect is observed in Fig. \ref{fig5}(a) for $\vv' = 0.6\,\hat{x}$, where we display a couple of skyrmions with total skyrmion charge $Q = 2$. 
The transition is originated due to the large self-induced field that results in an abrupt deformation of the skyrmion texture. Since changes in $Q$ characterizes the transition, the appearance of a new skyrmion is considered as a topological phase transition. Note that the pair of skyrmions at Fig. \ref{fig5}(a) does not return to the original state, characterized by $Q=1$, even when the velocity is set to zero. {Indeed, despite the $Q=1$ state has lower energy than states with $Q=2$, there is an energy cost to be overcome (topological protection) for the system back to the original state. Therefore, in the absence of any extra energy source, the system stays in the new state and no new topological transition are observed.}  In the limit of larger velocities, e.g., $\text{v}'=2$, the texture of skyrmions is destroyed, in agreement with results from Ref. [\onlinecite{LiuPRA}].  The increase in the skyrmion number for high current densities has been previously observed in Ref. [\onlinecite{Nagaosa-SciRep}], where the authors showed that at a current density threshold, the skyrmion suffers a substantial distortion, and consequently, a skyrmion multiplication is observed. Since the motion of skyrmions can be induced, for example, by electrical currents\cite{Schulz},
we estimate the necessary current density to induce the transition (shown at Fig. \ref{fig5}(a)) as $j \approx 1.64 \times10^{12}\ {\rm A}/{\rm m}^{2}$. This value is obtained considering $\Delta \rho_{xy}^{\infty}\approx 3\times 10^{-11}\ {\rm \Omega m}$ and $\Delta\rho_{xy}\approx4\times10^{-11}\ {\rm \Omega m}$ for Pd/Fe/Ir(111).

We now study the role of the system size and skyrmion velocity on the number of skyrmions appearing due to the topological transition. Because the skyrmion size is typically comparable to the length of the spiral determined by the competition among the DMI, exchange, anisotropy and Zeeman interactions \cite{Nagaosa,Nagaosa_Const}, one may expect that the quantity of skyrmions should be limited by the lattice size. Indeed, for a system with 15 sites per edge, a pair of skyrmions are stabilized, as shown in  Fig. \ref{fig5}(a). {However, for a system with 30 sites per edge and velocity $\text{v}' = 0.8$, the simulations reveal that $Q=9$ as shown in Fig. \ref{fig5}(b) and (c), after a total MCS of $N=N_d\,\times N_{\text{MCS}}$ with $N_d=5\times10^5$ and same initial conditions as the case presented in Fig. \ref{fig5}(a). Nevertheless, in addition to the lattice size, the skyrmion velocity also contributes to the total number of skyrmions. The relation between velocity and number of skyrmions can be determined from the total topological charge when the system reaches the steady-state. Fig. \ref{fig5}(c) depicts the value of $Q$ as a function of the skyrmion velocity for $N_{\text{MCS}}=120$. The inset depicts the evolution of the topological charge in terms of $N_\text{MCS}$ for various velocities. Note that $Q$ saturates for $\text{v}'\geq1.1$. This upper bound ($Q_u$) is due to the lattice size and, when 30 sites by edge is considered, we obtain that $Q_u=13$ which corresponds to the maximum number of skyrmions that can appear in the system. }

It is worth notice that although we assume that the spin-transfer torque effect induces the translational motion, our predictions are not restricted to electrical driving currents. Instead, it can be properly generalized to other mechanism like temperature gradients \cite{Yu988} and magnons-driven skyrmion motion \cite{GZhang-2018}.

\section{Conclusions}\label{Conclusions}
Using Monte Carlo simulations we study the influence of the motion of skyrmions on its shape in the range of low and high velocities. Our results shown that, depending on the velocity, a self-induced effective field ($\vB_{si}$) is generated and it induces changes in the skyrmion shape. 

In the low velocity regime we determined this effective field and used it to obtain the mass of skyrmions. For velocities above $\text{v}'=0.3$ (corresponding to $\text{v}= 27.4\ {\rm m/s}$), the field $\vB_{si}$ induces deformations on the skyrmion shape. The deformation breaks the circular symmetry and leads to an anisotropic mass. The effective mass $\cM_{xx}$ decreases and $\cM_{yy}$ increases as a function of the velocity. If the mechanism that enables the skyrmion movement is turned off at the low velocity regime, the skyrmion returns to its original circular symmetric shape. For very low velocities (i.e., lower than $\text{v}= 27.4\ {\rm m/s}$) the $|\vB_{si}|$ is small compared to the competing interactions, and the mass of the skyrmion remains almost constant. 

We have also shown that above a critical velocity, the self-induced field is large enough to modify the winding number. This characterizes a topological transition where the total number of new skyrmions is limited by the lattice size. {Additionally, we shown that the skyrmion velocity is crucial to determine the final topological charge of the system, where the number of skyrmions increase with the velocity. Nevertheless, there is an upper bound for the topological charge due to the lattice size, e.g., $Q_{u}=13$ for a system with 30 sites by edge.} We also confirmed that once the skyrmion velocity drops to zero, the system does not return to the original state, i.e., to a single skyrmion. Instead, the number of skyrmions remains constant. The mechanism behind the transition, e.g., the emergence of singularities during the skyrmion division \cite{Nature-2006,Thiaville-PRB,Gabriel-Work},  as well as the dynamics of the topological change are opened issues for future investigations. 

\section{Acknowledgments} In Brazil, this study was financed in part by the Coordena\c c\~ao de Aperfei\c coamento de Pessoal de N\'ivel Superior - Brasil (CAPES) -
Finance Code 001. The authors also thank CNPq (Grant Numbers 309484/2018-9, 306302/2018-7, and 302084/2019-3) and FAPEMIG for financial support. In Chile, we acknowledge support from Fondecyt grants 11170858 and 1200867
and Financiamiento Basal para Centros Cient\'{i}ficos y Tecnol\'ogicos de Excelencia AFB180001. R.E.T acknowledges the support by  the Research Council of Norway through is
Centres of Excellence funding scheme, Project No. 262633,
``QuSpin''.

\appendix

\end{document}